\documentclass{jpsj-suppl}
\usepackage{txfonts} 

\title{Power dependence of electric dipole spin resonance}

\author{Yasuhiro \textsc{Tokura}$^{1,2}$, Toshihiro \textsc{Kubo}$^{1}$ and William John \textsc{Munro}$^{2}$}

\inst{$^{1}$Graduate School of Pure and Applied Sciences, University of Tsukuba, Ibaraki 305-8571, Japan\\
$^{2}$NTT Basic Research Laboratories, NTT Corporation, Atsugi-shi, Kanagawa 243-0198, Japan}

\email{tokura.yasuhiro.ft@u.tsukuba.ac.jp}

\recdate{}

\abst{We develop a formalism of electric dipole spin resonance (EDSR) based on 
slanting magnetic field, where we especially investigate the microwave amplitude dependence.
With increasing microwave amplitude, the Rabi frequency increases linearly for
a spin confined in a harmonic potential.
However, when the spin is confined in the double-well potential, the Rabi
frequency shows sub-linear dependence with increasing the microwave amplitude.
}

\kword{electron spin, quantum dots, Rabi oscillations}

\begin{document}
\maketitle

\section{Introduction}

Coherent manipulation of the qubit is the most essential part of the quantum information processing.
In the solid state architecture of the quantum computing using electron spins in quantum dots (QDs)\cite{loss}, 
two-qubit operations are relatively easily realized and had been experimentally demonstrated 
since the accurate control of the spin exchange coupling energy $J$ is straightforward.\cite{petta}
In contrast, single spin coherent rotation is not easy.
Traditionally, spin manipulation is realized by electron spin resonance (ESR), where a time-dependent
transverse magnetic field of frequency $\omega $ close to $E_Z/\hbar$ is applied, with
$E_Z=g\mu_B B$ being the Zeeman energy by an external static magnetic field $B$, (with
the Land\'{e} g-factor, $g$, and the Bohr magneton, $\mu_B$).
Selective control of individual spins by ESR is hard, but the idea of electric dipole spin resonance (EDSR), 
which uses oscillating electric fields, instead of magnetic fields, had been proposed\cite{tokura,golovach} and
subsequently demonstrated\cite{michel,roland,nowack,nadj-perge,pribiag}.
The electron spin dipole is itself independent of the electric field, and the charge (orbital) degree of freedom 
is efficiently coupled to the electric field.
When there are some mechanisms to couple the orbital degree with the spin, 
the spin can be manipulated with electric field, which is called as EDSR.
The gradient of the static magnetic field\cite{tokura} and the spin-orbtit interaction\cite{golovach} 
are among these mechanisms. 

In the EDSR theory based on the magnetic field gradient\cite{tokura}, 
the effective Hamiltonian reads
\begin{eqnarray}\label{eq:rabi}
{\cal H}&=&\frac{E_Z}{2}\sigma_z+\frac{1}{2}\varepsilon_x\sigma_x\cos \omega t
\sim \frac{1}{2}(E_Z-\hbar\omega)\sigma_z+\frac{\varepsilon_x}{4}\sigma_x,
\end{eqnarray}
where $\varepsilon_x$ is the microwave-spin coupling constant and 
the last expression is obtained by the rotaing-wave approximation (RWA), 
assuming $\omega \gg |\varepsilon_x|/\hbar$.
At the resonant condition, $\omega=E_Z/\hbar$, the spin rotates around the $x$-axis of the Bloch sphere
with the Rabi frequency
\begin{eqnarray}
f_{Rabi}&=&\frac{\varepsilon_x(E_0)}{4\pi\hbar}.
\end{eqnarray}
The coupling constant $\varepsilon_x(E_0)$
was found linear with the amplitude of the applied electric field $E(t)=E_0\cos\omega t$
in the lowest order perturbation in $E_0$\cite{tokura}.
In particular for the harmonic potential, $V(z)=m\omega_0^2 z^2/2$, within the limit of small magnetic field, 
$E_Z\ll \hbar\omega_0$,
\begin{eqnarray}
\varepsilon_x(E_0)&\sim &\frac{g\mu_B b_{SL}\ell_0}{2\hbar\omega_0}(eE_0\ell_0),
\end{eqnarray}
where $m$ is the effective mass, $\ell_0\equiv \sqrt{\hbar/(m\omega_0)}$ and 
$b_{SL}$ characterizes the magnetic field gradient.
Although the result had been obtained by the perturbation approach,
it can be shown that the linearity of $\varepsilon_x$ to the field amplitude $E_0$ is 
exact for the harmonic potential\cite{tokura2} as far as RWA is legitimate.
It is interesting to investigate the field amplitude dependence of $\varepsilon_x$ 
for other types of confinement potentials.
Here, we studied this problem for the weakly coupled double QD and found clear non-linear behavior 
of the amplitude dependence of $\varepsilon_x$.

\section{Model}
The Hamiltonian is made of four terms
${\cal H}(t)={\cal H}_{spin}+{\cal H}_{orb}+{\cal H}_I+{\cal H}_{MW}(t),
$
where the first term is for the single spin under a static magnetic field $B$ in the $z$ direction
$
{\cal H}_{spin}=\frac{E_Z}{2}\sigma_z.
$
The second term is for a single electron orbital motion in a confinement potential $V(z)$,
$
{\cal H}_{orb}=\frac{p_z^2}{2m}+V(z),
$
where $p_z$ is the electron momentum operator.
For simplicity, we concentrate to the situation that the orbital motion is restricted in the 
one-dimensional system, and the motions toward the rest two directions are strongly suppressed.
The third term of the Hamiltonian is the coupling of the spin and orbital motion by
the slanting magnetic field,
$
{\cal H}_I=\frac{1}{2}g\mu_B b_{SL}\hat{z}\cdot \sigma_x,
$
where $b_{SL}$ characterizes the gradient in the $z$ direction of the magnetic field pointing to the $x$ direction.
Finally, the last term is the effect of microwave electric field with the frequency $\omega$,
$
{\cal H}_{MW}(t)=eE_0\hat{z}\cos\omega t,
$
where $E_0$ is the amplitude of the microwave.

We model the potential $V(z)$ as the double well-type, such that
\begin{eqnarray}
V(z)&=&\frac{m\omega_0^2}{8a^2}(z^2-a^2)^2,
\end{eqnarray}
where the length parameter $a$ characterizes the separation of the two minima.
When $\ell_0\equiv \sqrt{\hbar/(m\omega_0)}\ll a$, two gaussian localized wave functions at $z=\pm a$ are
relevant: $\varphi_{\pm a}(z)\equiv e^{-(z\mp a)^2/(2\ell_0^2)}/(\pi^{1/4}\sqrt{\ell_0})$.
We ortho-normalize these functions to 
\begin{eqnarray}
|L\rangle&\equiv &\frac{1}{\sqrt{1-2S g-g^2}}\{\varphi_{-a}(z)-g\varphi_{+a}(z)\},\\
|R\rangle&\equiv &\frac{1}{\sqrt{1-2S g-g^2}}\{\varphi_{+a}(z)-g\varphi_{-a}(z)\},
\end{eqnarray}
where $S\equiv \langle \varphi_{+a}|\varphi_{-a}\rangle$ and $g\equiv (1-\sqrt{1-S^2})/S$.
Now evaluating the matrix elements of ${\cal H}_{orb}$ using these basis functions,
$\varepsilon_d\equiv \langle L|{\cal H}_{orb}|L\rangle=\langle R|{\cal H}_{orb}|R\rangle$ and
$\Omega\equiv -\langle L|{\cal H}_{orb}|R\rangle=-\langle R|{\cal H}_{orb}|L\rangle>0$, we have the energy
eigenstates, 
$|S\rangle\equiv (|L\rangle+|R\rangle)/\sqrt{2}$ and $|A\rangle\equiv (|L\rangle-|R\rangle)/\sqrt{2}$ 
with eigenenergies, $E_S\equiv \varepsilon_d-\Omega$ and $E_A\equiv \varepsilon_d+\Omega$, respectively.
By using the Pauli matrices representing the psuedo-spin (orbital), we rewrite the orbital Hamiltonian as 
${\cal H}_{orb}=\Omega\tau_z$, omitting the constant energy $\varepsilon_d$.
Since the matrix elements of the position operator $\hat{z}$ are
$\langle S|\hat{z}|S\rangle=\langle A|\hat{z}|A\rangle=0$ and 
$\langle S|\hat{z}|A\rangle=\langle A|\hat{z}|S\rangle\equiv -Ca$, with
$C\equiv (1-g^2)/(1-2Sg-g^2)$, 
${\cal H}_{I}=\gamma \tau_x\sigma_x$ and ${\cal H}_{MW}(t)=\epsilon\tau_x\cos\omega t$ with
the spin-orbit coupling strength 
$\gamma\equiv -\frac{1}{2}g\mu_B b_{SL}Ca$ and 
the potential oscillation $\epsilon\equiv -eE_0 Ca$.
Hence, the total Hamiltonian reduces to
\begin{eqnarray}
{\cal H}(t)&=&\frac{E_Z}{2}\sigma_z+\Omega\tau_z+\gamma\tau_x\sigma_x+\epsilon\tau_x\cos\omega t,
\end{eqnarray}
and we need to solve the time-dependent Sch\"{o}dinger's equation 
$i\hbar\frac{\partial}{\partial t}\psi(t)={\cal H}(t)\psi(t)$.

\section{Result}
We introduce a canonical transformation with a Hermitian operator
$
{\cal A}\equiv \frac{\omega}{2}(\tau_z+\sigma_z),
$
such that for any operator ${\cal O}$,
$
\tilde{\cal O}(t)\equiv {\cal U}(t){\cal O}{\cal U}^\dagger(t),
$
where
$
{\cal U}(t)\equiv \exp({i{\cal A}t}).
$
Then, we define a new wave function
$
\phi(t)\equiv {\cal U}(t)\psi(t),
$
which satisfies following differential equation
\begin{eqnarray}
i\hbar \frac{\partial}{\partial t}\phi(t)&=&[-\hbar {\cal A}+\tilde{\cal H}(t)]\phi(t)
\equiv {\cal H}_{total}(t)\phi(t).
\end{eqnarray}
We then evaluate ${\cal H}_{total}(t)$.
Since $[{\cal A},\tau_z]=[{\cal A},\sigma_z]=0$, 
$
\tilde{\tau_z}(t)=\tau_z$ and $
\tilde{\sigma}_z(t)=\sigma_z.
$
It is easy to show following relations
$
\tilde{\tau}_x(t)=\tau_+e^{i\omega t}-\tau_-e^{-i\omega t},\ 
\tilde{\sigma}_x(t)=\sigma_+e^{i\omega t}-\sigma_-e^{-i\omega t},
$
where $\tau_{\pm}\equiv \frac{1}{2}(\tau_x\pm i\tau_y)$ and $\sigma_{\pm}\equiv \frac{1}{2}(\sigma_x\pm i\sigma_y)$.
Then we apply the rotating wave approximation (RWA) to ${\cal H}_{total}(t)$
where we neglect terms with $e^{\pm n \omega t}$, $n\ge 1$,
which would be a good approximation when $\gamma,\epsilon,\ll |\Omega-\frac{\hbar\omega}{2}|$.
Then we have
\begin{eqnarray}
{\cal H}_{RWA}&=&\frac{1}{2}(E_Z-\hbar\omega)\sigma_z+\Omega'\tau_z
+\frac{\epsilon}{2}(\tau_++\tau_-)+\frac{\gamma}{2}(\tau_+\sigma_-+\tau_-\sigma_+)
\end{eqnarray}
where we defined a reduced tunnel coupling energy 
$\Omega'\equiv \Omega-\frac{\hbar\omega}{2}\sim \Omega$ 
assuming 
$\Omega \gg \hbar\omega$.
We then diagonalize the orbital part
$
{\cal H}_{orb+MW}=\Omega'\tau_z+\frac{\epsilon}{2}(\tau_++\tau_-)$
into $W^{-1}{\cal H}_{orb+MW}W=\frac{1}{2}F\tau_z,
$
with $F\equiv \sqrt{(2\Omega')^2+\epsilon^2}$ by another unitary transformation  with
$W\equiv \hat{1}\cos\frac{\theta}{2}-i\tau_y\sin\frac{\theta}{2}$ and $\sin\theta\equiv \epsilon/F$.
The unitary operations applied to $\tau_x$ and $\tau_y$ are
$
W^{-1} \tau_x W=\tau_x\cos\theta+\tau_z\sin\theta,\ 
W^{-1} \tau_y W=\tau_y,
$
respectively.
Therefore, the total Hamiltonian reduces to
\begin{eqnarray}\label{eq:total}
\tilde{\cal H}_{RWA}&\equiv&W^{-1}{\cal H}_{RWA}W=\frac{F}{2}\tau_z
+\frac{1}{2}(E_Z-\hbar\omega)\sigma_z+\frac{\gamma\sin\theta}{2}\tau_z\sigma_x\nonumber\\
&+&\frac{\gamma}{2}\{(1+\cos\theta)(\tau_+\sigma_-+\tau_-\sigma_+)
+(1-\cos\theta)(\tau_-\sigma_-+\tau_+\sigma_+)\}.
\end{eqnarray}
For $F\gg \gamma, E_Z-\hbar\omega$, we can replace the operators related to the pseudo-spin (orbital) with its
expectation value, $\langle \tau_z\rangle \sim -1$ (the electron is assumed to be in the orbital ground state)
and the last term in Eq.(\ref{eq:total}), representing the real transitions in the orbitals, can be neglected,
otherwise, the coherence of the electron spin leaks to the orbital space, which results in spin decoherence.
In fact, this term is further removed by another unitary transformation with treating $\gamma$ as a perturbation,
which results in $\gamma^2/(F-E_z+\hbar\omega)$ order correction to the spin Zeeman energy.
Expanding $\sin\theta$ with assuming $\epsilon\ll \Omega'$, we have the effective Hamiltonian for the spin
\begin{eqnarray}
{\cal H}_{spin}&\sim &\frac{1}{2}(E_Z-\hbar\omega)\sigma_z
-\frac{\gamma \epsilon}{4\Omega'}\{1-\frac{1}{2}(\frac{\epsilon}{2\Omega'})^2\}\sigma_x.
\end{eqnarray}
Therefore, by comparing with Eq.(\ref{eq:rabi}), we have
\begin{eqnarray}
\varepsilon_x(E_0)&=&-\frac{g\mu_B b_{SL}Ca}{2\Omega-\hbar\omega}(eE_0Ca)
\{1-\frac{1}{2}(\frac{eE_0Ca}{2\Omega-\hbar\omega})^2\},
\end{eqnarray}
where $C\sim 1+S^2$ for $S\equiv \exp[-(\frac{a}{\ell_0})^2]\ll 1$.
Therefore the spin Rabi frequency proportional to $\varepsilon_x$ deviates from the linear in
$E_0$ dependence for larger microwave amplitude $E_0$.

\section{Conclusions}
We derived the Rabi frequency formula of the electron spin within the electric dipole spin resonance scheme, 
confined in a double quantum dot structure.
For sufficiently large microwave amplitude, the Rabi frequency deviates from the linear behavior with the amplitude.
However, for harmonic confinement potential, the Rabi frequency is expected to be linear with the microwave amplitude.
We anticipate the saturation of the Rabi frequency for an anhormonic confinement potential.
Recently, J. Yoneda {\it et al.}  achieved very large Rabi frequency as far as 100 MHz by optimizing the
design of the micro magnet and the quantum dot  as well as preparing large power of the microwave source\cite{yoneda}.
In the largest achievable  microwave amplitude regime, the Rabi frequency is deviated (saturated) from the 
linear dependence with the microwave amplitude\cite{yoneda2}.
Since the Rabi frequency is still much smaller than the applied microwave frequency, $\sim 10$GHz,
this saturation may not originate from simple factor of first-kind Bessel function\cite{nakamura}.
The obtained result in this work would be one of the possible mechanisms of this observation.
However, there may be another sources of the saturation of Rabi frequency, for example, 
real transitions to the excited orbital states, the non-linearity of the slanting field, and Joule heating.
These are the subjects for future study.

We thank J. Yoneda and S. Tarucha for useful discussions. 
Part of this work is supported by JSPS MEXT Grant-in-Aid for Scientific Research 
on Innovative Areas (21102003) and Funding Program for 
World-Leading Innovative R\&D Science and Technology (FIRST).

\end{document}